\documentclass[preprint,aps,superscriptaddress,showpacs,showkeys,nofootinbib,notitlepage ]{revtex4-1}
\usepackage{graphicx}
\usepackage{amsmath}
\usepackage{longtable}
\usepackage{color}

\newcommand{\be}{\begin{equation}}
\newcommand{\ee}{\end{equation}}
\newcommand{\bea}{\begin{eqnarray}}
\newcommand{\eea}{\end{eqnarray}}
\newcommand{\etal}{et al.}

\newcommand{\pht}{\tilde{\phi}_1}
\newcommand{\bd}{\beta_d}
\newcommand{\bdelt}{\mbox{\boldmath$\delta$}}

\def\sarc{$^{\prime\prime}\!\!.$}

\def\marc{$^{\prime}\!.$}

\begin{document}

\bibliographystyle{apsrev}

\title{Fermat's least-time principle and the embedded transparent lens}

\author{R. Kantowski}
\email{kantowski@ou.edu}
\affiliation{Homer L.~Dodge Department~of  Physics and Astronomy, University of
Oklahoma, 440 West Brooks,  Norman, OK 73019, USA}

\author{B. Chen}
\email{bchen@ou.edu}
\affiliation{Homer L.~Dodge Department~of  Physics and Astronomy, University of
Oklahoma, 440 West Brooks,  Norman, OK 73019, USA}

\author{X. Dai}
\email{xdai@ou.edu}
\affiliation{Homer L.~Dodge Department~of  Physics and Astronomy, University of
Oklahoma, 440 West Brooks,   Norman, OK 73019, USA}
\date{\today}

\begin{abstract}
We present a simplified version of the lowest-order embedded point mass gravitational lens theory and then make the extension of this theory to any embedded transparent lens.
Embedding a lens effectively reduces the gravitational potential's range, i.e., partially shields the lensing potential because the lens mass is made a contributor to the mean mass density of the universe and not simply superimposed upon it.
We give the time-delay function for the embedded point mass lens from which  we can derive the simplified lens equation by applying Fermat's least-time principle.
Even though rigorous derivations are only made for the point mass in a flat background, the generalization of the lens equation to lowest-order for any distributed lens in any homogeneous background is obvious.
We find from this simplified theory that embedding can introduce corrections above the few percent level in weak lensing shears caused by large clusters but only at large impacts. The potential part of the time delay is also affected in strong lensing at the few percent level.
Additionally we again confirm that the presence of a cosmological constant alters the gravitational deflection of passing photons.
\end{abstract}

\pacs{98.62.Sb}
\keywords{General Relativity; Cosmology; Gravitational Lensing;}
\maketitle
\section{Introduction}

In this paper we present a simplified version of the embedded lens theory given in Refs.\,\cite{Kantowski10,Kantowski12,Chen10,Chen11} which can easily be used by anyone familiar with conventional lensing theory, see Sec.\,\ref{sec:Lowest}.
Our investigations into embedded lenses began as an attempt to settle a dispute about the effect, or the lack thereof, of the cosmological constant $\Lambda$ on gravitational lensing \cite{Rindler07,Khriplovich08,Park08,Sereno09,Schucker09,Simpson10,Ishak10,Bhadra10,Arakida12}. To be certain that the usual lensing approximations would not introduce erroneous $\Lambda$ effects, we found it necessary to strictly impose Einstein's gravity theory while making the lensing mass a contributor to the mean mass density of the universe. We found that such a lens, an embedded lens, was effectively shielded beyond a certain range and that shielding has a far more important effect on time delays and weak lensing shears than does the presence of the cosmological constant.
Our simplified lens theory will significantly simplify modeling effects of individual inhomogeneities (e.g., clusters of galaxies and cosmic voids) on observed cosmic microwave background (CMB) temperature anisotropies at large angles, in particular modeling the integrated Sachs-Wolfe (ISW) effect \cite{Sachs67}.

Unfortunately the theory, in the form previously presented and to an accuracy required to detect $\Lambda$ corrections, is sufficiently complicated to discourage its use. However, we have now
succeeded in simplifying the theory by replacing previously used image impact position variables by observable image position angles, and by keeping only the most important terms, i.e., correction terms caused by shielding. In this paper we present this theory in a form which is easily compared with the conventional theory.
We start with a short description of the embedded lens model and the perturbation scheme we use to compute the relevant lensing quantities. We then describe how the current simplified results are arrived at from previous more complicated expressions by changing the independent lensing variables from the minimum radial coordinate $r_0$ and the local impact angle $\pht$, see Fig.\,1, to the observer's image angle $\theta_I$, see Fig.\,2.
In addition to giving the lens equation as a function of $\theta_I$ we also give the time-delay function,  i.e., the excess time light takes to reach the observer because it encounters the embedded point mass lens.  By applying Fermat's least time principle \cite{Schneider85,Blandford86} to this time-delay function the entire lens theory, to the lowest order, is reproduced.
We also conclude that because we can interpret the lowest-order equation within the conventional theory as lensing by a point mass plus a negative surface mass density,
we are able to extend this theory to any embedded transparent lens, see Sec.\,\ref{sec:Extending}.

\section{The Embedded Lens Model}\label{sec:Embedded}

\begin{figure*}
\includegraphics[width=0.8\textwidth,height=0.26\textheight]{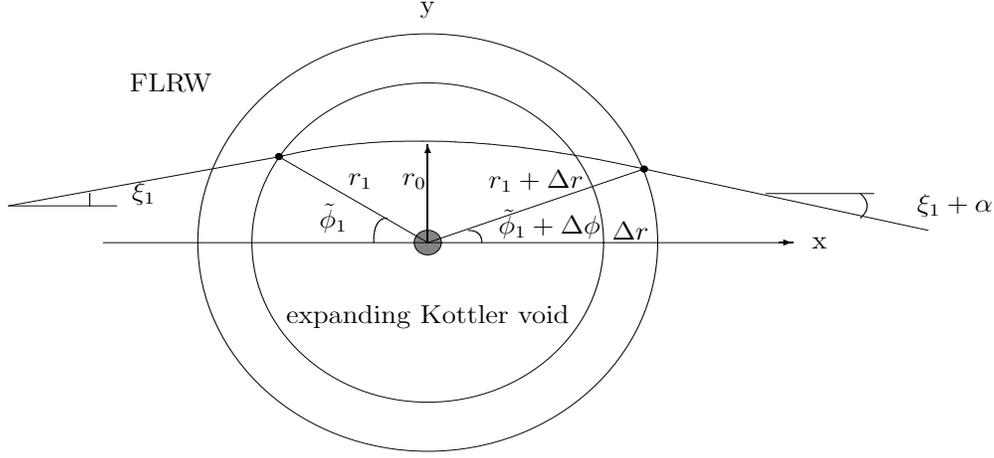}
\caption{A photon travels left to right entering a Kottler void at $r=r_1,\, \phi=\pi-\tilde{\phi}_1$
and returns  to the FLRW dust at $r=r_1+\Delta r,\, \phi=\tilde{\phi}_1 +\Delta\phi$.
The photon's orbit has been chosen symmetric in Kottler coordinates about the point of closest approach $r=r_0$, $\phi=\pi/2$. Due to the cosmological expansion, $\Delta r>0$.
The slope of the photon's co-moving trajectory in the x-y plane is $\xi_1$
when incoming and $\xi_1+\alpha$ after exiting. The deflection angle $\alpha$, see Eq.\,(\ref{alpha}), is negative by convention and is precisely and uniquely defined by this figure.
Expressions for $r_1,$ $\Delta r,$ $\xi_1,$ and $\Delta\phi$ as functions of the two impact parameters, $r_0$ and $\pht$, can be found in Refs.\,\cite{Kantowski10,Chen10,Chen11}. The minimum radial coordinate $r_0$ as a function of $\pht$ was given in Eq.\,(A1) of Ref.\,\cite{Kantowski12}  and allowed $r_1$, $\alpha$, etc., to be given as functions of $\pht$ alone.}
\label{fig:fig1}
\end{figure*}

In previous work Refs.\,\cite{Kantowski10,Kantowski12,Chen10,Chen11} we have investigated, in depth, effects of embedding on the point mass lens equation and on all resulting image properties.
However, application of that theory is quite complex because of the lens impact variables used.
Our embedded point mass lens remains a spherical comoving mass void  (see Figs.\,1 and 2) in a flat pressureless Friedman-Lema\^itre-Robertson-Walker (FLRW) universe, containing the cosmic mass $m$ removed from the void condensed at its center \cite{Einstein45,Schucking54}.
The actual geometry in the void is described by the Kottler metric \cite{Kottler18} which is similar to Schwarzschild's geometry but with the addition of a cosmological constant $\Lambda$.
This lens model is used to maintain absolute correctness of the gravity theory and hence avoid approximation errors in the lensing theory. A universe filled with such voids is called a Swiss cheese universe and has historically been used to compute effects of inhomogeneities on distance-redshift relations \cite{Kantowski69,Dyer74,Schneider88,Kantowski95,Kolb10,Clarkson12,Fleury13} and on the anisotropy of the CMB \cite{Rees68,Dyer76,Martinez90,Sakai08,Granett08,Ade13}.
The comoving radius $\chi_b$ of a lens void centered at comoving distance $\chi_d$, equivalent to  a redshift $z_d$  from the observer in the FLRW background (see Fig.\,2), is related to the Schwarzschild radius $r_s\equiv2G{\rm m}/c^2$ of the embedded lens by
 \be
 r_s=\Omega_{\rm m}\frac{H_0^2}{\rm c^2}\chi_b^3,
 \label{rs}
 \ee
where $H_0$ is the Hubble constant, $\Omega_{\rm m}$ is the usual matter density parameter, and the current radius of the flat universe has been taken equal to 1. Throughout, a subscript `$d$' on evolving quantities means evaluated precisely at the cosmic time $t_d$ equivalent to  $z_d$. We refer to $z_d$ as the deflector's redshift, and think of $t_d$ as the time when the observed photons passed the deflector, even though a passing time is not precisely defined. For example the lens void expands with the background cosmology and had a physical radius $r_d=\chi_b/(1+z_d)$ at the precise time $t_d$.

In the next section we explain the approximation procedure
we use  to obtain the embedded point mass lensing results and in Sec.\,\ref{sec:Changing} we arrive at the new lensing formulae from the previous more complicated expressions. The key is to use the image position angle as the independent variable rather than an impact variable. In Sec.\,\ref{sec:Lowest} we drop all but the lowest order terms and obtain what we call the simplified embedded theory.  In Sec.\,\ref{sec:Extending} we then explain how the simplified, but rigorous, point mass results are generalized to include distributed lenses.
\begin{figure*}
\includegraphics[width=1.0\textwidth,height=0.19\textheight]{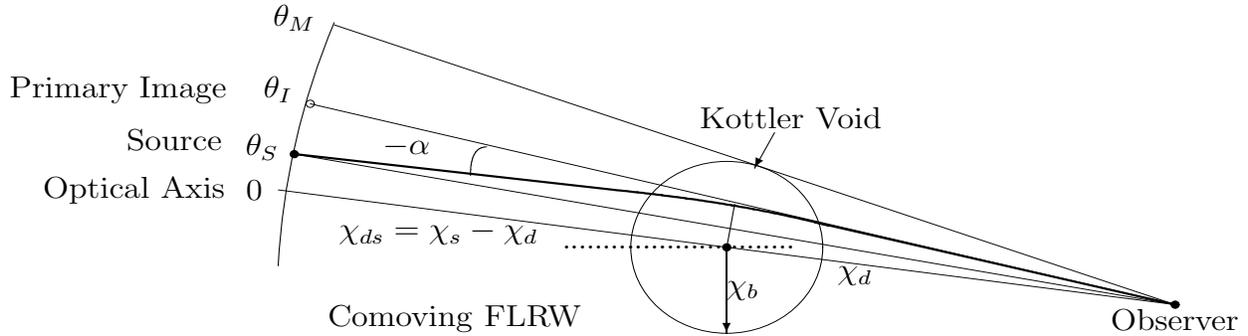}
\caption{ Lensing geometry of an embedded point mass lens.
A photon travels from a source located at angular position $\theta_S$, a comoving distance $\chi_s$ from the observer, and enters a Kottler void of comoving radius $\chi_b$ centered at comoving distance $\chi_d$ from the observer.
The photon is deflected by an angle $\alpha$ ($<0$) and returns  to the FLRW dust on its way to the observer, where it appears at angle $\theta_I$.
The maximum image angle $\theta_I$ is $\theta_M\equiv\chi_b/\chi_d$ and occurs when the light ray grazes the Kottler void. The significant simplification of the lens theory given in this paper occur because we have been able to replace  $\pht$ of Fig.\,1 with the image angle $\theta_I$ shown here by using Eq.\,(\ref{pht}).}
\label{fig:fig2}
\end{figure*}
\section{The Approximation Technique}\label{sec:Approximation}
Because embedded lensing quantities we wish to compute, such as the bending angle, the image magnification, etc., depend nonlinearly on small parameters we resort to a series expansion of the desired quantities in these small parameters. For an embedded point mass lens calculation in a flat FLRW background cosmology  where the image angle $\theta_I$ is used as the independent variable, three independent small parameters are relevant: $\chi_b/\chi_d,\ r_s/r_d,$ and  $\Lambda r_d^2,$  (see Figs.\,1 and 2). Rather than introducing three independent perturbation parameters, one for each of these small quantities, we observe from Eq.\,(\ref{rs}) that for a fixed FLRW cosmology and fixed lens redshift $z_d$,
the three parameters vary with the embedded mass as $\chi_b/\chi_d\propto r_s^{1/3},\ r_s/r_d\propto r_s^{2/3},$ and  $\Lambda r_d^2\propto r_s^{2/3}.$
The first of the three parameters $\chi_b/\chi_d$ is the same as the angular radius $\theta_M$ of the void as seen by the observer, see Fig.\,2.
When we are interested in a galaxy or cluster lens at $z_d=0.5$ in the concordance cosmology $\theta_M$ is of order $10^{-2}$--$10^{-3}$. The next two parameters are approximately of a numerical order that is the square of that. Consequently we can keep track of the various orders of approximation by introducing a single perturbation parameter $\bdelt$. If we want to keep only first-order terms in a calculation, $\bdelt^1$ terms, we only keep terms containing $\theta_M$ to the first power. If we want to  keep second-order terms, $\bdelt^2$ terms, we additionally keep $(\theta_M)^2, \ r_s/r_d,$ and  $\Lambda r_d^2$ terms. Extension to higher order terms is obvious. To actually isolate terms of the same order in an expression we multiply each by the appropriate power of $\bdelt$ and collect all terms of that same power. As is common practice with perturbation techniques, we then give $\bdelt$ the numerical value of 1 so as not to alter the value of the expression.

For the embedded point mass lens an additional complication occurs. The lowest-order terms explicitly containing $ r_s/r_d,$ and  $\Lambda r_d^2$ are square roots of the form
\be
\bd\equiv \frac{v_d}{c}=
\sqrt{r_s/r_d+\Lambda r_d^2/3}=\frac{H_d}{c}r_d,
\label{beta}
\ee
where $v_d$ is the expansion rate  of the void's boundary as measured by  static observers at the boundary
and where $H_d$ is the Hubble parameter at the lens redshift $z_d.$
Equation (\ref{beta}), as well as Eq.\,(\ref{rs}), are consequences of the geometry of embedding. When $\bd$ appears in an expression it is given the order $\bdelt^1$ and is kept as $\bd$ for notational convenience even though it is not an independent parameter.

If the reader wants to apply our results to a lens where the three parameters $\theta_M,\ r_s/r_d,$ and $\Lambda r_d^2$  are not of the respective orders discussed above he/she can do so. The results given will be correct but the reader simply has to be aware that  all terms collected as $\bdelt^n$ terms for a fixed $n$ are not necessarily  numerically similar.

The approximation technique described here was used in Ref.\,\cite{Kantowski12} where a local impact angle $\pht$ was used rather than the image position angle $\theta_I$. In Refs.\,\cite{Kantowski10,Chen10,Chen11} we used the minimum impact coordinate $r_0$ rather than $r_d$ to form two of the three small expansion parameters; however, the expansion technique was essentially the same.

\section{Changing Independent Variables}\label{sec:Changing}

The rigorously derived embedded point mass lens equation to order $\bdelt^4$ was given in our earlier work Refs.\,\cite{Kantowski10,Chen10} where we computed lensing quantities as functions of two  impact variables of the Kottler metric, the minimum impact radius $r_0$ and the void impact angle $\pht$ , see Fig.\,1.
In Refs.\,\cite{Chen11,Kantowski12} we succeeded in eliminating the minimum impact radius $r_0$ in favor of the single void impact angle $\pht$, see Eq.\,(A1) of Ref.\,\cite{Kantowski12}, and were consequently able to write the lens equation, etc., as functions of $\pht$ alone.
Only now have we succeeded in eliminating $\pht$ in favor of the observation angle $\theta_I$ and in truly simplifying the embedded point mass lens equation.

The steps we follow to obtain the simplified  theory are as follows:
By iteratively inverting the image position $\theta_I(\pht)$ as given in Eq.\,(8) of Ref.\,\cite{Kantowski12} as a function of previously used impact variable $\pht$, we obtain $\pht$ as a function of $\theta_I/\theta_M$
\be
\pht = \sin^{-1}(\theta_I/\theta_M)+\bdelt \bd\ (\theta_I/\theta_M)
-\bdelt^2\left\{\frac{\theta_M^2}{6}\frac{(\theta_I/\theta_M)^3}{\sqrt{1-(\theta_I/\theta_M)^2}}
+\frac{r_s}{r_d}\frac{\left[\sqrt{1-(\theta_I/\theta_M)^2}\right]^3}{(\theta_I/\theta_M)}\right\}+{\cal O} \bigl(\bdelt^3\bigr).
\label{pht}
\ee
In this and in the following  expressions $\theta_I\le\theta_M$ are both individually treated as order $\bdelt^1$ but $\theta_I/\theta_M\le 1$ is of order $\bdelt^0$.
For the orbit approximation used to derive the embedded lens theory to be valid the following lower limits on orbit parameters must be adhered to: $\sin\pht\ll r_s/r_0\Rightarrow\sin^2\pht\ll r_s/r_d\Rightarrow (\theta_I/\theta_M)^2\ll r_s/r_d$.
We could give Eq.\,(\ref{pht}) to a higher order but for our current purposes it isn't needed.
By using Eq.\,(\ref{pht}) to eliminate $\pht$  in the deflection angle $\alpha(\pht)$ given by Eq.\,(7) of Ref.\,\cite{Kantowski12} the embedded point mass deflection angle shown in Figs.\,1 \& 2 is written as a function of the image angle $\theta_I$ as
\bea\label{alpha}
\alpha &=&-\frac{2r_s}{r_d}\Biggl[\Biggl[
 \frac{\left[\sqrt{1-(\theta_I/\theta_M)^2}\right]^3}{(\theta_I/\theta_M)}\nonumber\\
&+&\bdelt^2\,\Biggl\{
\theta_I\sqrt{1-(\theta_I/\theta_M)^2}
\left(
\frac{1}{6} \theta_M \left[1+2(\theta_I/\theta_M)^2\right]
-\frac{1}{2}\beta_d\left[1-(\theta_I/\theta_M)^2\right]
\right)\nonumber\\
&+&\frac{r_s}{r_d}
\Biggl(
\frac{1}{16}\frac{\sqrt{1-(\theta_I/\theta_M)^2}}{(\theta_I/\theta_M)}
\left[3-14(\theta_I/\theta_M)^2+20(\theta_I/\theta_M)^4 \right]\nonumber\\
&+&\frac{15}{16}\frac{\cos^{-1}\left[\theta_I/\theta_M\right]}{(\theta_I/\theta_M)^2}-
\frac{3}{4}(\theta_I/\theta_M)\log\left[\frac{1+\sqrt{1-(\theta_I/\theta_M)^2}}{1-\sqrt{1-(\theta_I/\theta_M)^2}}\right]
\Biggr)\Biggr\}+{\cal O} \bigl(\bdelt^3\bigr)
\Biggr]\Biggr],
\eea
and from the lens equation (6) of Ref.\,\cite{Kantowski12}, the embedded lens equation now becomes
\bea \label{lens-eqn}
\theta_S&=&\theta_I -\frac{\theta_E^2}{\theta_M}
\Biggl[\!\Biggl[
 \frac{\left[\sqrt{1-(\theta_I/\theta_M)^2}\right]^3}{(\theta_I/\theta_M)}
+\bdelt^2\Biggl\{\ \frac{r_s}{r_d}\Biggl[\frac{15}{16}\frac{\cos^{-1}\left(\theta_I/\theta_M\right)}{(\theta_I/\theta_M)^2}\nonumber\\
&&+\frac{1}{16}\frac{\sqrt{1-(\theta_I/\theta_M)^2}}{(\theta_I/\theta_M)}
\left[3-14(\theta_I/\theta_M)^2+20(\theta_I/\theta_M)^4\right]
 -\frac{3}{4}(\theta_I/\theta_M)\log\frac{1+\sqrt{1-(\theta_I/\theta_M)^2}}{1-\sqrt{1-(\theta_I/\theta_M)^2}}\Biggr]\nonumber\\
&-&\frac{1}{2}\beta_d\,\theta_M\Biggl[(\theta_I/\theta_M)\sqrt{1-(\theta_I/\theta_M)^2}\left(1-(\theta_I/\theta_M)^2-\frac{2}{3}\frac{\chi_d}{\chi_{ds}}\left[4-(\theta_I/\theta_M)^2\right]\right)\nonumber\\
&&+\frac{\chi_d}{\chi_{ds}}(\theta_I/\theta_M)\log\frac{1+\sqrt{1-(\theta_I/\theta_M)^2}}{1-\sqrt{1-(\theta_I/\theta_M)^2}}\Biggr]
\\
&+&\frac{1}{6}\theta_M^2\Biggl[(\theta_I/\theta_M) \sqrt{1-(\theta_I/\theta_M)^2}
\left(
  \left[1+2(\theta_I/\theta_M)^2\right]+3\frac{\chi_d}{\chi_{ds}}\left[1-(\theta_I/\theta_M)^2\right]\right)\Biggr]
\Biggr\}+{\cal O} \bigl(\bdelt^3\bigr)
\Biggr]\!\Biggr]\nonumber.
\eea
The only new parameter  appearing in the embedded point mass lens equation is the familiar non-embedded angular Einstein ring radius
\be
\theta_E\equiv \sqrt{\frac{2\ r_s D_{ds}}{D_d D_s}},
\ee
where $D_d,$ $D_s,$ and $D_{ds}$ are angular diameter distances, respectively of the deflector, source, and source relative to the deflector.
The embedded value of the Einstein ring size is found by putting $\theta_S=0$ and iteratively solving for $\theta_I$.
The maximum value of the image angle $\theta_I$ is $\theta_M$ and occurs for the primary image when the light ray from the source just grazes the spherical void.
For this maximum image angle, the deflection angle is zero and the source is likewise located at $\theta_M$.
For a massive galaxy or cluster lens ($M=10^{12} M_\odot$ or $10^{15} M_\odot$) at redshift $z_d=0.5,$ and a source at $z_s=1.0,$ the Einstein ring radius $\theta_E$ is respectively 1\sarc66 or 52\sarc6,  much smaller than the maximum image angle, i.e., $\theta_E/\theta_M =0.0085$ or 0.027. Consequently both strong and weak lensing effects are seen within the void.

The lowest-order dependence of the bending angle $\alpha$ on $\Lambda$ is clearly demonstrated by the presence of $\bd$ in Eq.\,(\ref{alpha}) at second-order.
The vanishing of terms of order $\bdelt^1$  is surprising since they are  present when
$\pht$ is used as the independent image variable, see Eqs.\,(6) and (7)  of Ref.\,\cite{Kantowski12}.
What we see is that when $\theta_I$ is used as the independent variable, terms of order $\bdelt^1$ all vanish, including $\bd$ terms, and hence $\Lambda$ is no longer present at first-order \cite{Schucker10,Boudjemaa11}.
This seems somewhat paradoxical, $\Lambda$ does/doesn't affect bending at a given order depending on what independent variable is used, $r_0$ or $\theta_I$. But as with most paradoxes the controversy is one of definitions; the $\bdelt^0$ term in Eq.\,(\ref{alpha}) will again produce a $\bdelt^1$ term when the inverse of Eq.\,(\ref{pht}) is used to replace $\theta_I$ by $\pht$.

The initial controversy over  $\Lambda$'s effect on lensing seems to persist primarily  because  photon orbits in the Kottler space-time, when parameterized by $r_0$, do not depend on $\Lambda$. The lack of an effect on the photon's orbit is extended beyond the point mass lens to include distributed mass lenses by working in appropriate weak field gauges.  The orbit's independence of $\Lambda$, along with the verbal justification that $\Lambda$ acts isotropically, has been enough to continue fueling the no-$\Lambda$  effect supporters \cite{Khriplovich08,Park08,Simpson10,Arakida12}.
Rindler and Ishak \cite{Rindler07}  have  given the most significant criticism of the no-$\Lambda$  effect position by pointing out that simply knowing an orbit is not sufficient to measure an angle; to do that you additionally need an observer and a reference direction. Obviously the value of the angle measured depends on which observer is doing the measuring and what reference direction is used \cite{Sereno09,Schucker09,Ishak10} but all reasonable choices produce  bending angles that depends on $\Lambda$ at some order. For our embedded lens  we have made the only logical choice of observers, comoving FLRW observers, and the only logical reference direction, the initial direction of the photon, for measuring the deflection angle $\alpha$ (see Fig.\,1). The observers are unique, and because the spatial FLRW background is flat, the initial photon direction is defined for all comoving observers and all times. Because the bending angle is determined by the gravitational interaction occurring over a small spatial extent (i.e., while in the void) cosmological distances cannot be intrinsic contributors to $\alpha$'s dependence on $\Lambda$.  For example, $\alpha$ as given by Eq.\,(32) of Ref.\,\cite{Kantowski10} doesn't contain any dependence on $\Lambda$ via a cosmological distance.
However, by using orbit parameters and impact variables that depend on angular diameter distances such additional $\Lambda$ dependence can be introduced as seen in Eq.\,(7) of Ref.\,\cite{Kantowski12} and  Eq.\,(\ref{alpha}) above. At what order that dependence appears depends on which  independent orbit variable is used, $\pht$ or $\theta_I$.

The embedded lensing time-delay function $T(\theta_S,\theta_I)$ as given next is the difference in arrival times of  two signals, one lensed and one not, both starting simultaneously at a fixed comoving source distance $\chi_s$ and reaching the observer at respective times $t_0$ and $t_0-T(\theta_S,\theta_I)$.
The total time delay is commonly written as a sum of  geometrical and  potential parts \cite{Cooke75,Chen10}, which for the embedded point mass, were given  in Equations (10) and (14) of Ref.\,\cite{Chen10} and Equations (22) and (23) of Ref.\,\cite{Kantowski12} as functions of impact variables $r_0$ and $\pht$.
By eliminating $r_0$ and $\pht$ in favor of $\theta_I$ and
by using the lens equation to simplify the geometrical part (the first part containing $(\theta_S-\theta_I)^2/2$)
we obtain
\bea
cT(\theta_S,\theta_I)&=& (1+z_d)\frac{D_d D_s}{D_{ds}}\Biggl\{\frac{(\theta_S-\theta_I)^2}{2}
-\theta_E^2\Biggl[\frac{1}{2}\log\left(\frac{\theta_I/\theta_M}{1+\sqrt{1-(\theta_I/\theta_M)^2}}\right)^2\nonumber\\
& &+\frac{1}{3}\left[4-(\theta_I/\theta_M)^2\right]\sqrt{1-(\theta_I/\theta_M)^2}+{\cal O} \bigl(\bdelt^2\bigr)\Biggr]\Biggr\}.
\label{Fermat}
\eea
This expression, without the $(1+z_d)$ factor, is the embedded {\it Fermat potential} \cite{Schneider92} whose minimization gives the lens equation \cite{Schneider85,Blandford86}, but only accurate to order $\bdelt^1$, compare Eqs.\,(\ref{lens-eqn}) and (\ref{lens-eqn1}).
In Eq.\,(\ref{Fermat}) we have given $T(\theta_S,\theta_I)$ to the maximum accuracy we can, i.e., to order $\bdelt^1$, given that we have computed the lens equation to an accuracy no higher than $\bdelt^4$.
In other words we would need the lens equation to even higher order than Eq.\,(\ref{lens-eqn}) to compute the {\it Fermat potential} to any higher order.

\section{The Embedded Lens Theory to Lowest Order}\label{sec:Lowest}

If we drop the $\bdelt^2$ terms in Eq.\,(\ref{lens-eqn}) we obtain what we call the simplified embedded point mass lens equation
\be
\theta_S=\theta_I-\frac{\theta_E^2}{\theta_I}\left[\sqrt{1-(\theta_I/\theta_M)^2}\right]^3.
\label{lens-eqn1}
\ee
It is accurate to order $\bdelt^1$ even though all $\bdelt^1$ terms happen to vanish because $\theta_I$ is used as the independent variable.
This equation is also obtained by varying $cT(\theta_S,\theta_I)$ of Eq.\,(\ref{Fermat}) with respect to the image angle $\theta_I$, i.e., by applying {\it Fermat}'s least time principle.
The conventional linear point mass lensing equation, without embedding, is regained by neglecting the $(\theta_I/\theta_M)$ term.

Image properties can easily be derived from
 Eq.\,(\ref{lens-eqn1}).  The inverse image matrix eigenvalues (respectively axial and radial) are found to be
\bea\label{aphi}
a_\phi&=&1-\left(\frac{\theta_E}{\theta_I}\right)^2\left[\sqrt{1-(\theta_I/\theta_M)^2}\right]^3,\cr
a_r&=&1+\left(\frac{\theta_E}{\theta_I}\right)^2\left[1+2\left(\theta_I/\theta_M\right)^2 \right]\sqrt{1-(\theta_I/\theta_M)^2}.
\eea
The reciprocal amplification is
\be
\mu^{-1}
=1+3\left(\frac{\theta_E}{\theta_M}\right)^2\sqrt{1-(\theta_I/\theta_M)^2}-\left(\frac{\theta_E}{\theta_I}\right)^4\left[1+2\left(\theta_I/\theta_M\right)^2 \right]\left[1-(\theta_I/\theta_M)^2\right]^2,
\ee
with a normalized effective surface mass density $\kappa$ present
\be
\kappa
=-\frac{3}{2}\left(\frac{\theta_E}{\theta_M}\right)^2\sqrt{1-(\theta_I/\theta_M)^2}.
\label{kappa}
\ee
This angular dependent negative term exactly equals the projected surface mass density of the removed homogeneous sphere at the image position $\theta_I.$
The shear in the bundle of light rays at image position $\theta_I$ is
\be
\gamma
= \frac{1}{2}\left(\frac{\theta_E}{\theta_I}\right)^2[2+(\theta_I/\theta_M)^2]\sqrt{1-(\theta_I/\theta_M)^2}.
\label{gamma}
\ee
By neglecting the $\theta_E/\theta_M$ and $\theta_I/\theta_M$ terms, i.e., the shielding terms caused by embedding, the above expressions all reduce to the conventional non-embedded values for the point mass lens.
These results represent a significant simplification in our previous work on embedded lenses and can easily be used without following the complexities of their derivation outlined in the previous section.
\begin{figure*}
\begin{center}$
\begin{array}{cc}
\includegraphics[width=0.5\textwidth,height=0.32\textheight]{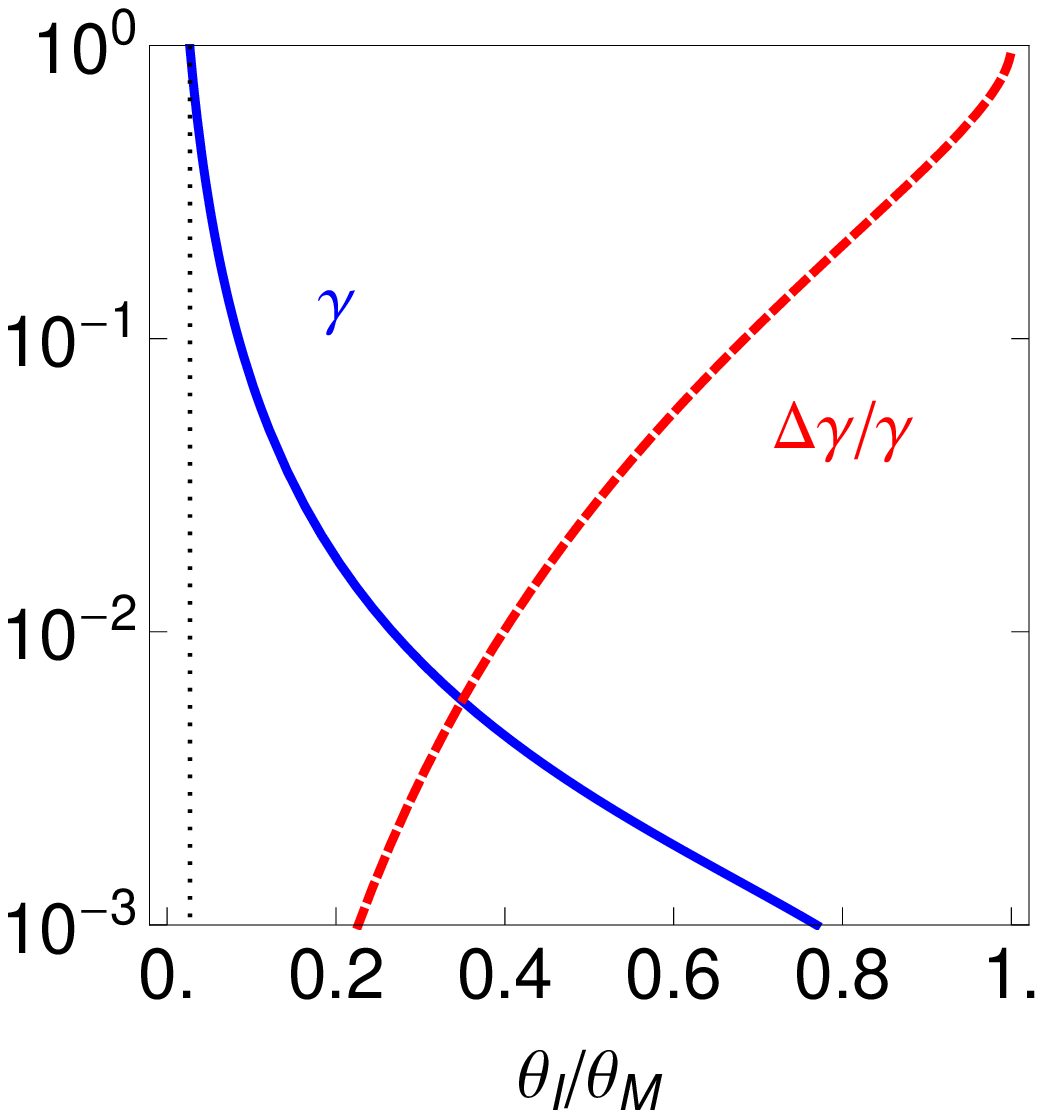}
\hspace{15pt}
\includegraphics[width=0.5\textwidth,height=0.32\textheight]{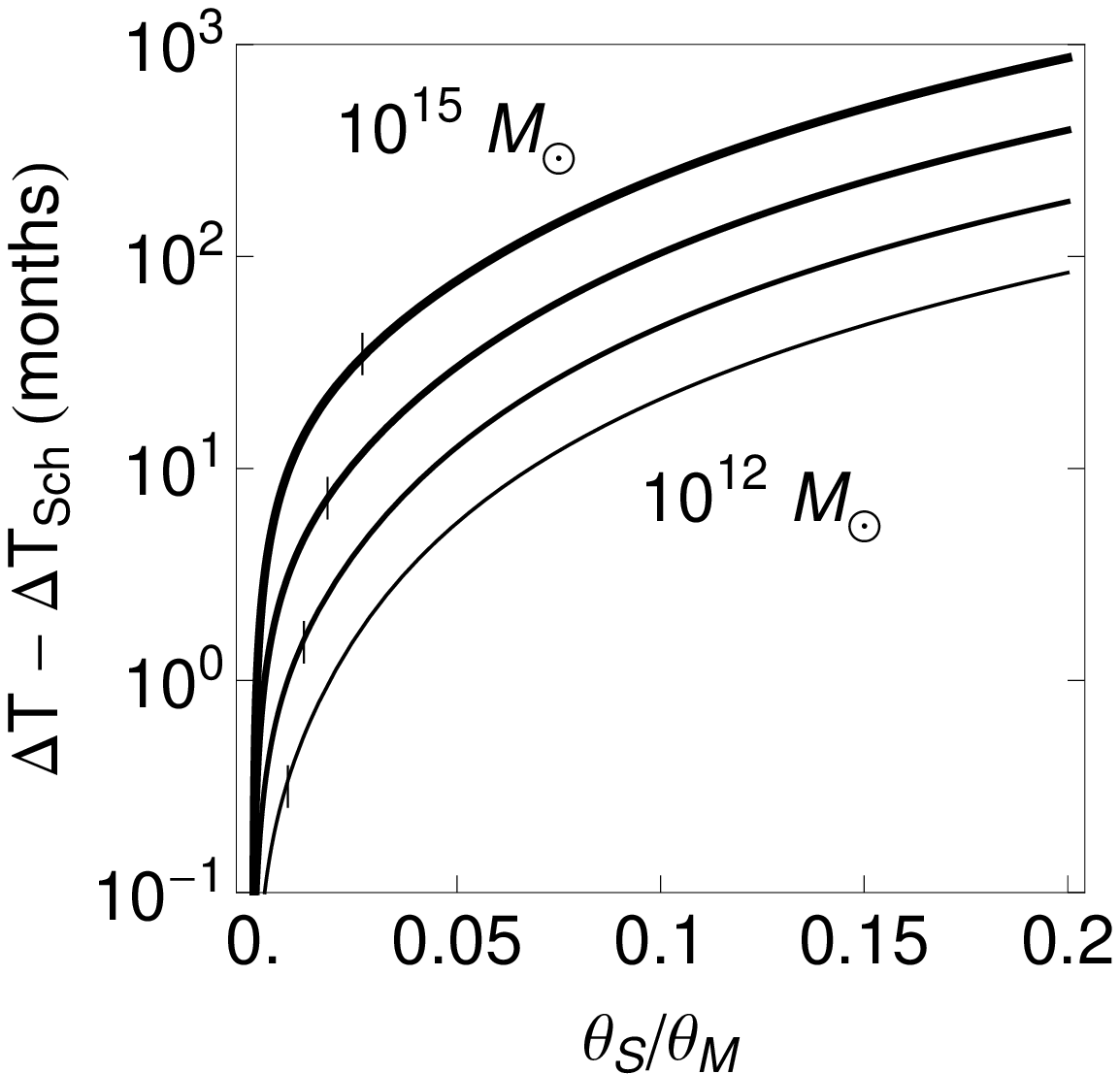}
\end{array}$
\end{center}
\caption{ (Left) Embedding corrections to the image shear $\gamma$ for a $10^{15} M_\odot$ cluster lens.
The source and lens redshifts are  $z_s=1.0$ and $z_d=0.5,$ respectively.
The solid blue curve  is the shear, $\gamma$,  and the dashed red curve  is the fractional difference in the image shear, $\Delta\gamma/\gamma$, caused by embedding. Both are plotted as a functions of $\theta_I/\theta_M$ where the Einstein ring radius is $\theta_E= $52\sarc6, the void radius is $\theta_M= $32\marc8, and $\theta_E/\theta_M=0.027$ is the vertical dotted line.
(Right) The difference in the embedded time delay and the conventional Schwarzschild time delay, $\Delta T-\Delta T_{\rm Sch}$,  computed using Eq.\,(\ref{Fermat}) and the Schwarzschild equivalent for the primary and secondary images, is plotted as a function of source position angle $\theta_S/\theta_M$. The time delay difference is given in months for four point mass lenses, $ m=10^{15}M_\odot,10^{14}M_\odot,10^{13}M_\odot,$ and $10^{12}M_\odot$ (respectively top to bottom curves). The small vertical ticks on each curve are at respective $\theta_E/\theta_M$ values.
 }
\label{fig:fig3}
\end{figure*}
In Fig.\,3  on the left we have plotted the shear $\gamma$ as a function of $\theta_I$ and have compared the embedded $\gamma$ with the conventional value for a large cluster lens. The shear is seen to differ from the conventional Schwarzschild value by more than a few percent but only at impact angles more than $15\,\theta_E$, i.e., in the weak lensing regime.
In Fig.\,3 on the right we have plotted  time delay differences caused by embedding. These differences are primarily due to the potential part of Eq.\,(\ref{Fermat}) differing from the Schwarzschild value at large primary image angles.
The generalization of the point mass lens equation
to spherically distributed and arbitrarily distributed lenses is given in the next section.

\section{Extending The Embedded Lens Theory To Distributed Lenses}\label{sec:Extending}

The point mass theory is of limited usefulness on the scale of clusters.
Fortunately the above simplified theory is easily generalized to lenses with arbitrarily distributed mass, once it is observed that the order $\bdelt^1$ lens equation (\ref{lens-eqn}) is the same as the classical lens equation for the superposition of a point mass and a distributed mass lens with negative projected surface mass density $\Sigma(\theta_I)\equiv\Sigma_{\rm cr}\kappa(\theta_I),$ where
 $\kappa$ is given in Eq.\,(\ref{kappa}) and $\Sigma_{\rm cr}\equiv c^2D_s/4\pi GD_dD_{ds}$ is the conventional critical surface mass density \cite{Schneider92}.
This leads to the immediate observation that we can generalize, to this order of accuracy, lens equation (\ref{lens-eqn}) to include any transparent spherically symmetric mass
\be
\theta_S=\theta_I-\frac{\theta_E^2}{\theta_I}\left[f(\theta_I)-\left(1-\left[\sqrt{1-(\theta_I/\theta_M)^2}\right]^3\right)\right],
\label{spherical-lens}
\ee
where $f(\theta_I)$ is the fraction of the actual distributed mass lens within the impact cylinder defined by  $\theta_I$ (equivalently the fraction projected within the impact disc defined by $\theta_I$).
Because the exact Einstein models are known, the Lema\^itre-Tolman-Bondi metrics \cite{Lemaitre33,Tolman34,Bondi47} embedded in FLRW,  this result can be confirmed by rigorously extending the work done in Refs.\,\cite{Kantowski10,Kantowski12,Chen10,Chen11}.
The generalization to any distributed lens, confined within the lens but not necessarily spherically symmetric,  is also obvious
\be
\boldsymbol{\theta}_S=\boldsymbol{\theta}_I
+\frac{D_{ds}}{D_s}\boldsymbol{\alpha}_0(\boldsymbol{\theta}_I)
+\boldsymbol{\theta}_I\left(\frac{\theta_E}{|\boldsymbol{\theta}_I|}\right)^2\left(1-\left[\sqrt{1-(|\boldsymbol{\theta}_I|/\theta_M)^2}\right]^3\right),
\label{arbitrary-lens}
\ee
where $\boldsymbol{\alpha}_0$ is the conventional bending angle without embedding.
This result cannot be readily confirmed because the relevant Einstein solutions are not known.
Nonetheless, it should be correct to order $\bdelt^1$.
Image properties for these transparent lenses can be obtained much as they were for the point mass in Eqs.\,(\ref{aphi})--(\ref{gamma}).
Because of the simple dependence of these lowest order embedded lens equations on angular diameter distances, they should also be valid for any FLRW background models,  not just spatially flat cosmologies.

The simplified and generalized transparent lens theory presented in this paper is expected to be useful for studies of weak lensing by galaxy clusters, cosmic shears, and lensing of the CMB radiation by large scale structures.
An investigation of the late time ISW effect caused by embedded galaxy clusters and cosmic voids using the formalism developed in this paper is in preparation.

\begin{acknowledgments}

B.C. acknowledges the following support: NSF AST-0707704, and US DOE Grant DE-FG02-07ER41517 and Support for Program number HST-GO-12298.05-A was provided by NASA through a grant from the Space Telescope Science Institute, which is operated by the Association of Universities for Research in Astronomy, Incorporated, under NASA contract NAS5-26555.
\end{acknowledgments}

\end{document}